\documentclass[a4paper,12pt]{article}
\usepackage{epsfig}
\usepackage[T1]{fontenc}
\usepackage{graphicx}
\usepackage{amsfonts}
\usepackage{latexsym}
\usepackage{amsmath}
\usepackage{subfigure}
\usepackage{mathrsfs}

\newcommand{\ham}{{\mathcal H}}

\begin{document}

\vskip 3truecm
\begin{verbatim}
\end{verbatim}
\bigskip\bigskip\bigskip\bigskip\bigskip

\centerline{\Large\bf Dynamics of the two-dimensional gonihedric spin model } 
\bigskip
\begin{center}
{\sc D. Espriu\footnote{espriu@ecm.ub.es} 
and A. Prats\footnote{aleix@ecm.ub.es}}\\
{\em Departament d'Estructura i Constituents de la
Mat\`eria}\\
{\em Universitat de Barcelona, Diagonal 647, 08028 Barcelona, Spain}\\
\end{center}

\bigskip\bigskip

\begin{abstract}
In this paper we study dynamical aspects of the two-dimensional (2D)
gonihedric spin model using both numerical and analytical
methods. This spin model has vanishing microscopic surface tension and
it actually describes an ensemble of loops living on a 2D surface. The
self-avoidance of loops is parameterized by a parameter $\kappa$. The
$\kappa=0$ model can be mapped to one of the six-vertex models
discussed by Baxter and it does not have critical behavior. We have
found that $\kappa\neq 0$ does not lead to critical behavior
either. Finite size effects are rather severe, and in order to
understand these effects a finite volume calculation for non
self-avoiding loops is presented. This model, like his 3D counterpart,
exhibits very slow dynamics, but a careful analysis of dynamical
observables reveals non-glassy evolution (unlike its 3D
counterpart). We find, also in this $\kappa=0$ case, the law that
governs the long-time, low-temperature evolution of the system,
through a dual description in terms of defects. A power, rather than
logarithmic, law for the approach to equilibrium has been found.
\end{abstract}

\vfill

UB-ECM-PF-03/38

December 2003\\

\eject

\section{Introduction}
The gonihedric spin model that we are going to study in this paper
 was first introduced by Savvidy in higher dimensions as 
a discretized model for
tensionless string theory \cite{Amba-suka,Savv-al.}. Very
soon the spin model gained interest by itself in its three dimensional
version. Also its extension to self-interacting surfaces (parameterized
by a self-avoidance parameter $\kappa$) led to a family of
models with different critical behavior and interesting dynamical
properties. Extensive numerical and theoretical work appeared
\cite{Savv,Kout-savv.bis,Kout-savv,Ba-es-jo-ma,John-malm},
showing that the behavior of this 3D model turns out to
be glassy \cite{Es-lip-jo,Swi-bo-tra-ba,Lip,Lip-jo,Dim-es-ja-pra} even if
no disorder is present. The 2D version of the model turns
out to have trivial thermodynamics  but rather peculiar dynamical
properties and this is the reason that motivated us to investigate
this model in detail. To our knowledge, the only existing study of
 this 2D version is some work \cite{Bu-gar} related to the
 fluctuation-dissipation theorem. It has been suggested that an
 experimental realization of this type of models (see \textit{e.g.}
 \cite{Savv}) could be of application to magnetic memory devices. 
 
The Hamiltonian for the gonihedric spin model adapted to a 2D
 embedding space is the following
\begin{equation}
\ham_{gonih}^{\textit{\tiny 2D}}=
-\kappa\sum_{<i,j>}\sigma_i\sigma_j 
+\frac{\kappa}{2}\sum_{\ll i,j\gg}\sigma_i\sigma_j
-\frac{1-\kappa}{2}\sum_{[i,j,k,l]}\sigma_i\sigma_j\sigma_k\sigma_l 
\nonumber 
\end{equation}
where $\sigma_i$ are spin variables on the sites of a 2D
square lattice, $<i,j>$ means sum over nearest neighbor pairs, $\ll
i,j\gg$ means sum over next to nearest neighbor pairs, and
$[i,j,k,l]$ means sum over groups of four spins forming elementary
plaquettes in the lattice. The coupling
constants of this model are very finely tuned. The dynamics of models
with nearest-neighbor and next-to-nearest-neighbor interactions only
have been studied elsewhere \cite{Sho-hol-set}. A competing 
nearest- and next-to-nearest-neighbor model is obtained for 
$\kappa=1$,  where the plaquette term is absent, but it
turns out that the gonihedric $\kappa=1$ case lies
just outside the parameter space they analyzed. The
geometric interpretation is missing in the choice of couplings of
\cite{Sho-hol-set}. 

The energy landscape of this family of models is very peculiar in any
number of dimensions, due mainly to the large amount of symmetry of
the ground state. 
This symmetry consists, in the 2 dimensional case, in the possibility
of flipping all spins contained in one row or one column of the
lattice without changing the energy of the ground-state\footnote{There is actually a difference in the symmetry operations
  you can perform in the $\kappa=0$ and the $\kappa\neq 0$ case. In the
  first case you can flip a row or a column of spins without any
  constrain. In the second case, from a ferromagnetic state you can
  flip either only columns or only rows. Flipping one set of spins of
  each type increases the energy due to the generation energetic
  configurations at the meeting point of the row and column.}. 
This symmetry reveals a huge degeneracy of the ground
state. This fact together with the dynamically generated energy
barriers\footnote{In the 2 dimensional case the barriers that the
  model generates dynamically are not dependent of the length of the
  domain (unlike the 3D version) and this will make a difference in
  the dynamical behavior.} that the system encounters then cooling
down provides the ingredients to exhibit glassy behavior, and the
3-dimensional model indeed does exhibit that type of behavior
\cite{Lip-jo,Dim-es-ja-pra}.  

\begin{figure}[!ht]
\begin{center}
\includegraphics[scale=0.6]{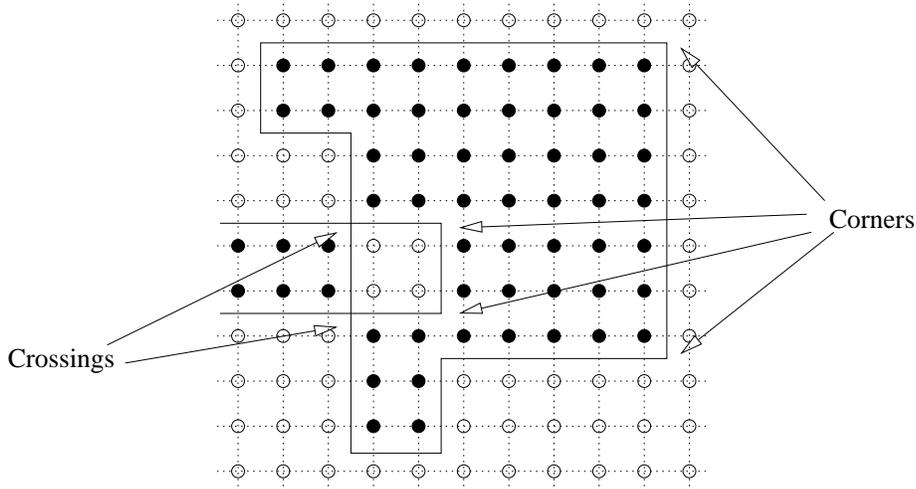}
\end{center}
\caption{Example of the correspondence between spins and loops. All
  the energy is concentrated in the corners and crossings of the loops.}  
\label{2Dexample}
\end{figure}

The $\kappa$ parameter regulates the self-avoidance of the interface
(lines in 2 dimensions) of up and down spins. We focus our attention
on the properties of the interface between up and down spins, because
in the bulk we know that there is no excess energy. As can be seen in
fig.\ref{2Dexample} these interface form loops that may have
self-crossings.  

By looking at the energy of the loop model it is not difficult to see
that it can be written as
$E=n_2+4\kappa n_4$, where $n_2$ is the number of bending points of the
loops formed by the interface, and $n_4$ is the number of self-crossings.
That means that $\kappa\to 0$ is the limit for the non-self-avoiding
loops, and the $\kappa\to\infty$ is the completely-self-avoiding limit,
in which no crossing of loops can exist. 
Thus the system likes flat interfaces. This is the main reason for the
creation of energy barriers while cooling. The system tends to flatten
its interface at a first stage, but this process favors 
configurations where square domains of any size appear, and at low
temperature those configurations are very stable.  

In the next section we review the main thermodynamical
features of the model. Section \ref{dynamics} is dedicated to a
numerical study of the dynamics of this models at low
temperature in order to determine whether there is glassy behavior in the 2
dimensional gonihedric model as it is actually the case in the 3
dimensional one. In Section \ref{longtimes} we carry out an
analytical study of the behavior of the system at low temperatures
and long times that we then proceed to compare to a numerical
analysis. Our conclusions are collected in section \ref{conclusions}.
We relegate  some technical details to two appendices.

\begin{figure}[!ht]
\begin{center}
\subfigure[]{\includegraphics[scale=0.29,angle=-90]{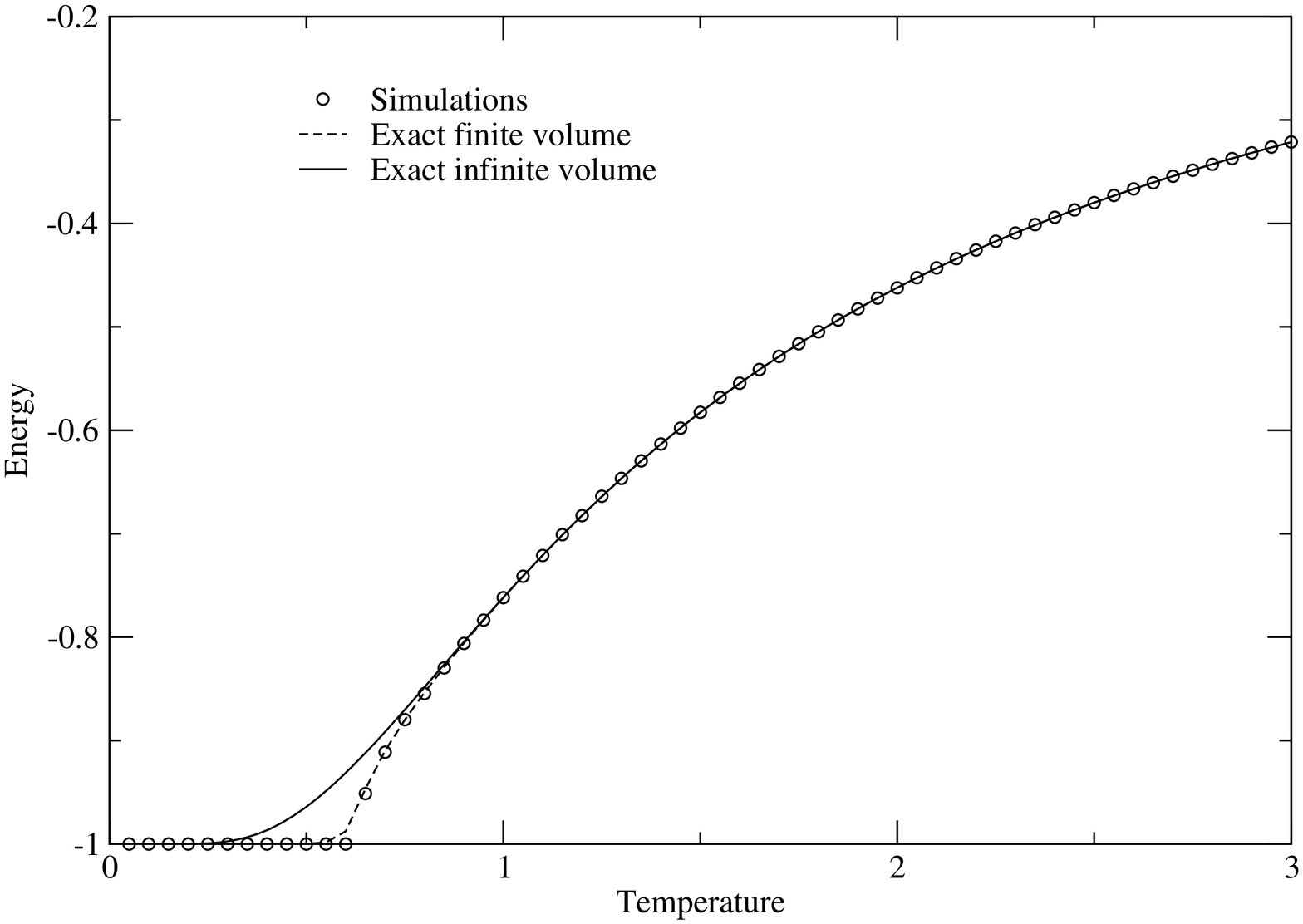}}
\subfigure[]{\includegraphics[scale=0.29,angle=-90]{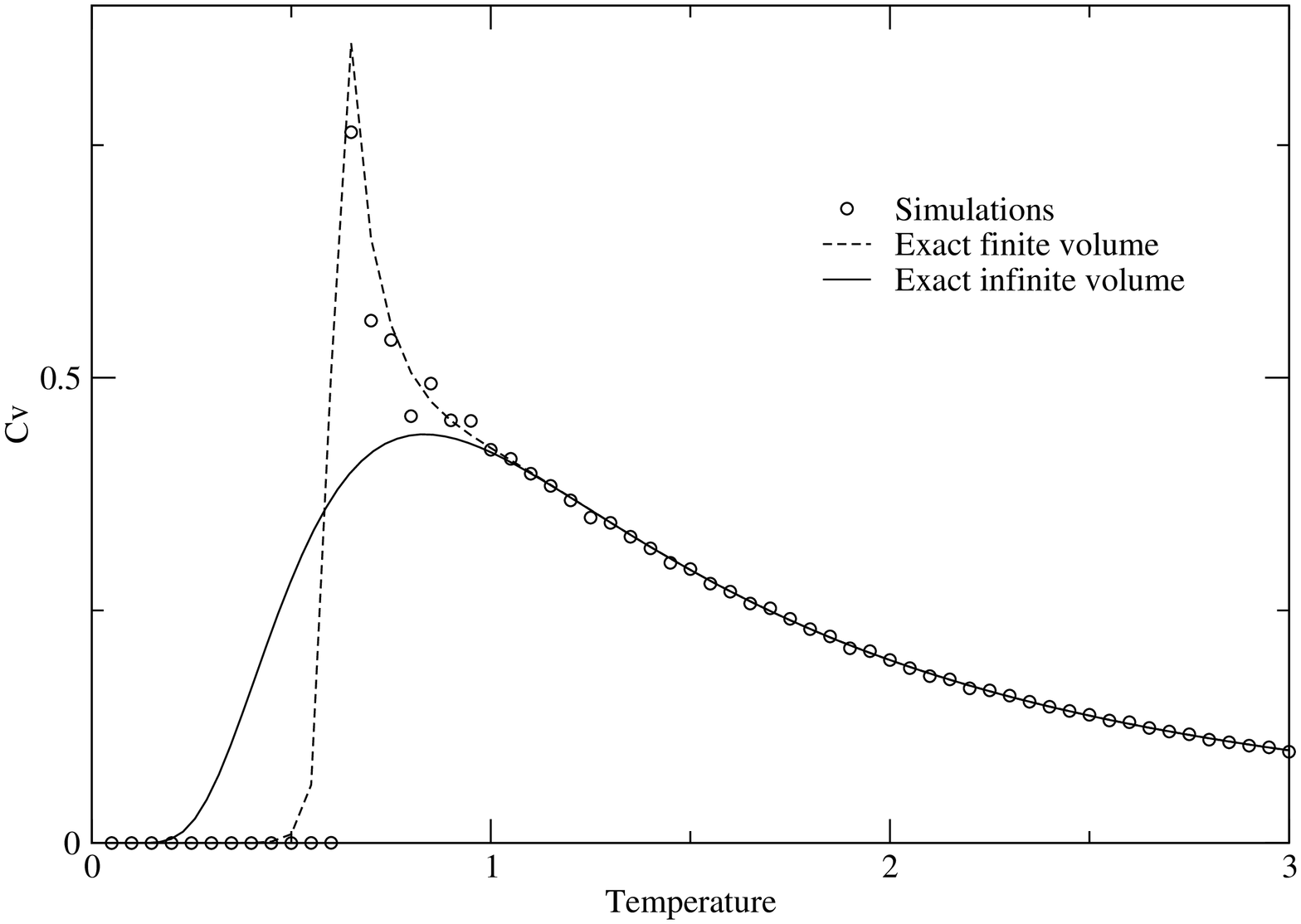}}
\end{center}
\caption{(a) Energy function and (b) specific heat of the system for
  $\kappa=0$. The exact function at infinite volume, at finite volume,
  and the Monte-Carlo simulations are plotted.}
\label{EneK0}
\end{figure}
\section{Thermodynamics of the model}\label{thermodynamics}

Let us begin with the simplest case $\kappa=0 $ that is
exactly solvable in the infinite volume limit and can be 
reduced to an easy-computable sum for finite volume (see appendix A). 
The exact solution for the model with $\kappa=0$ \cite{Baxter} shows
that there is no phase transition at finite temperature. If we take
a look at fig.\ref{EneK0} we will see the infinite volume energy
function and specific heat compared to the numerical results and to
the exact finite volume calculation. All discrepancies between
simulations and the infinite volume calculations are due to finite
volume effects as we can see comparing the simulations with the exact
finite volume calculations\footnote{It is clear that in this model the
  finite volume effects are very important, mostly around the
  temperature where the maximum in the specific heat is placed. The
  finite volume calculations are performed with a $100^2$ volume and
  periodic boundary conditions.}.  
\begin{figure}[!ht]
\begin{center}
\subfigure[]{\includegraphics[scale=0.29,angle=-90]{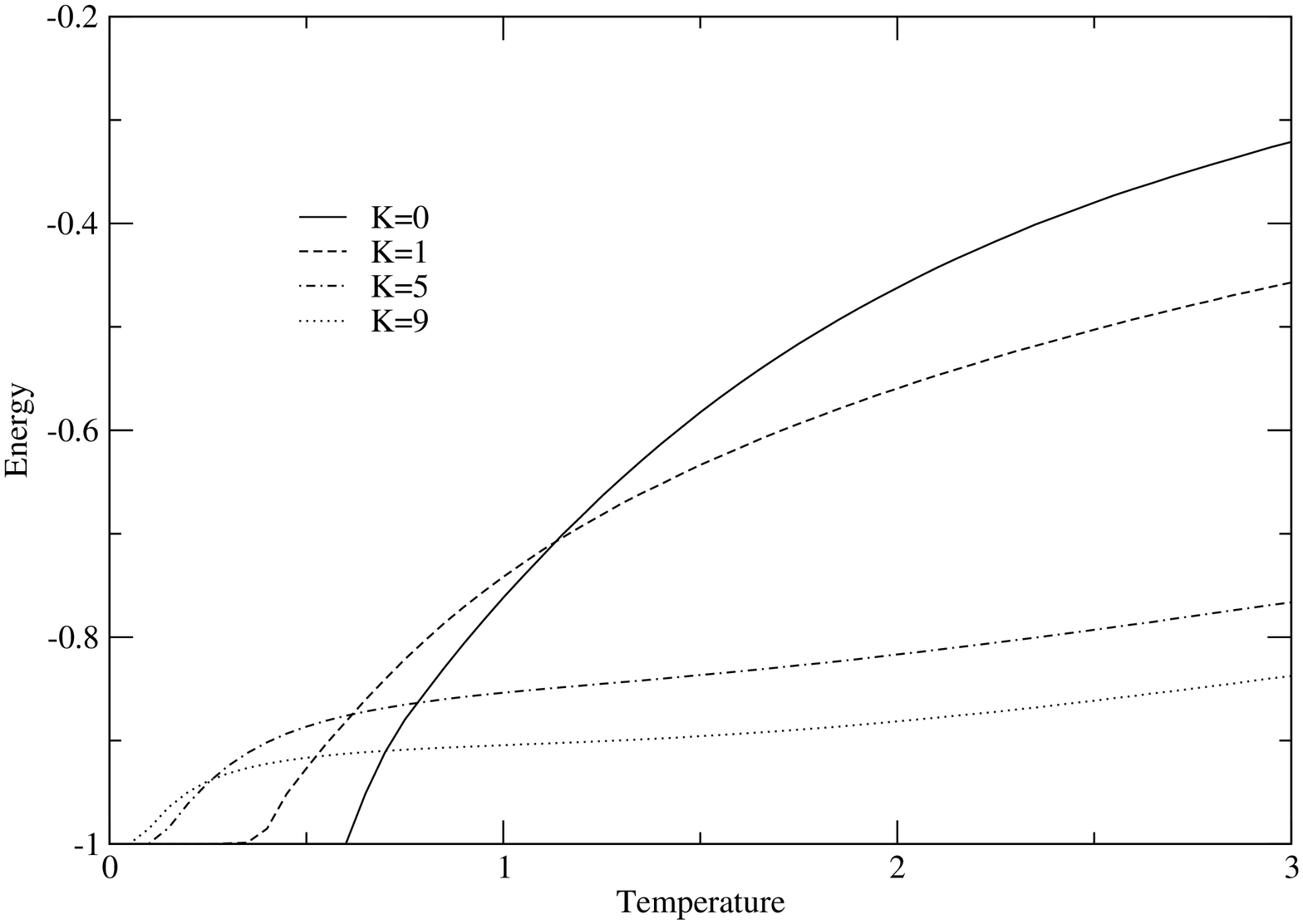}}
\subfigure[]{\includegraphics[scale=0.29,angle=-90]{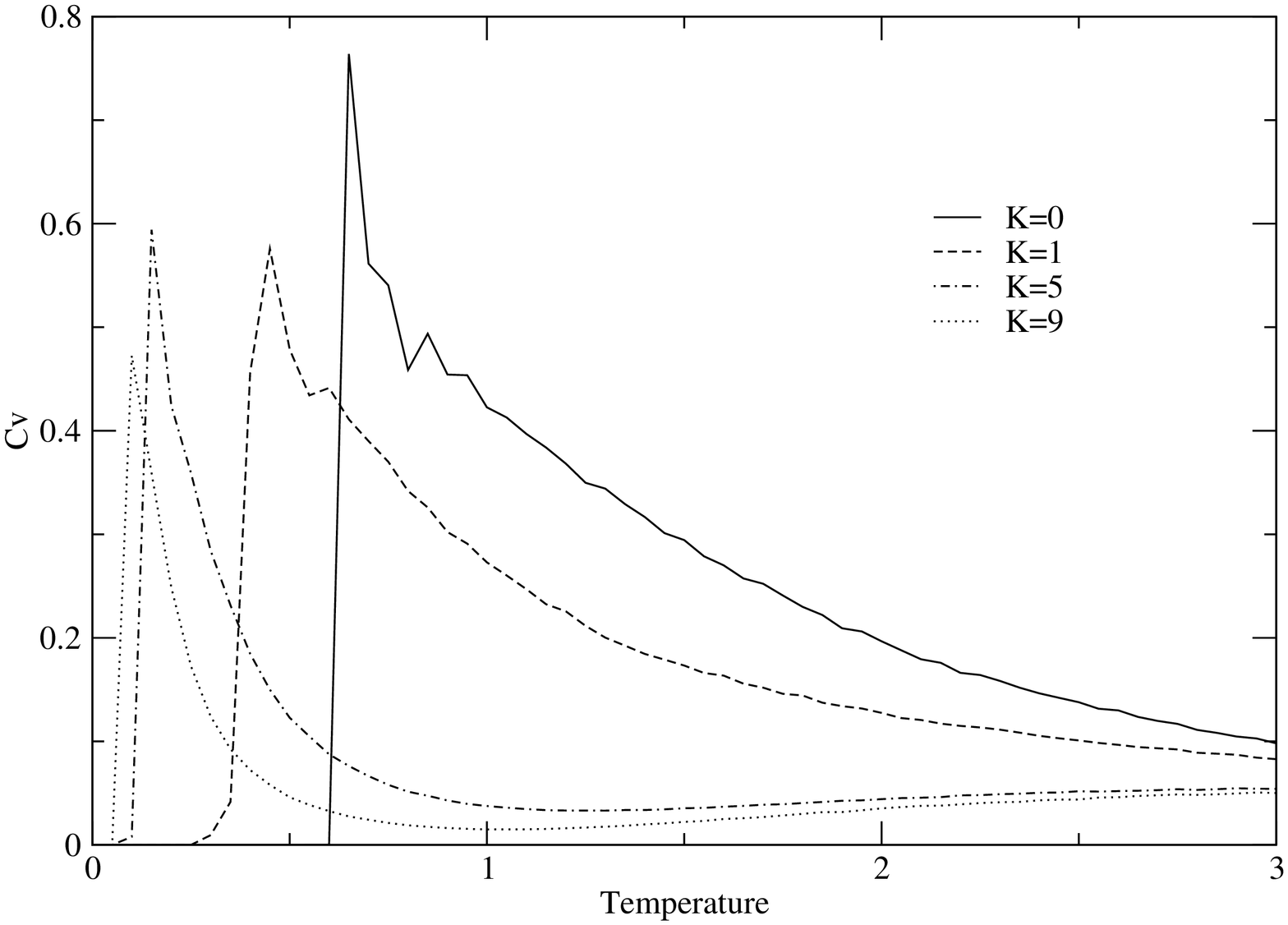}}
\end{center}
\caption{(a) Energy function and (b) specific heat of the system for
  different values of $\kappa$. Only simulations are plotted. Signals
  of the non-monotonic behavior can be seen for the $\kappa=5$ and
  $\kappa=9$ case in the high temperature region of the plot.}  
\label{EneKno0}
\end{figure}
For the other case with $\kappa\ne 0$ there is no exact infinite 
volume solution or easy-computable finite-volume expression, but the
simulations do not show marked differences with the $\kappa
=0$ case, so we are forced to conclude that there is no ordered phase
at low temperature (see fig.\ref{EneKno0})\footnote{Can be seen from
  this plots that the energy has been rescaled in order to have
  energy $-1$ at zero temperature. The same kind of convenient
  rescaling of the temperature and the specific heat with a factor
  depending only on $\kappa$ has been performed to compare the
  different simulations.}.   
The maximum in the specific heat seen in the $\kappa=0$ case is still
present at the same point (as it should, because
it reveals the temperature where the first excitations appear in the
bulk) and behaves in the same way. The only remarkable difference is the
appearance of a second structure for sufficiently large $\kappa$
(An indication for this can be seen in
fig.\ref{EneKno0}b in the non-monotonic behavior of the specific heat
for $\kappa=5$ and $9$. Notice the rescaling of the data mentioned in
the footnote).   
This second structure can be interpreted as the appearance of a new
state for the plaquette variables whose energy grows with $\kappa$.
No volume dependence of this structure has been found, so there is no
evidence of phase transition. In figures \ref{secondarystructure}a and
\ref{secondarystructure}b we can actually see the formation of this secondary 
structure and its independence on the volume, respectively.  
\begin{figure}[!ht]
\begin{center}
\subfigure[]{\includegraphics[scale=0.29,angle=-90]{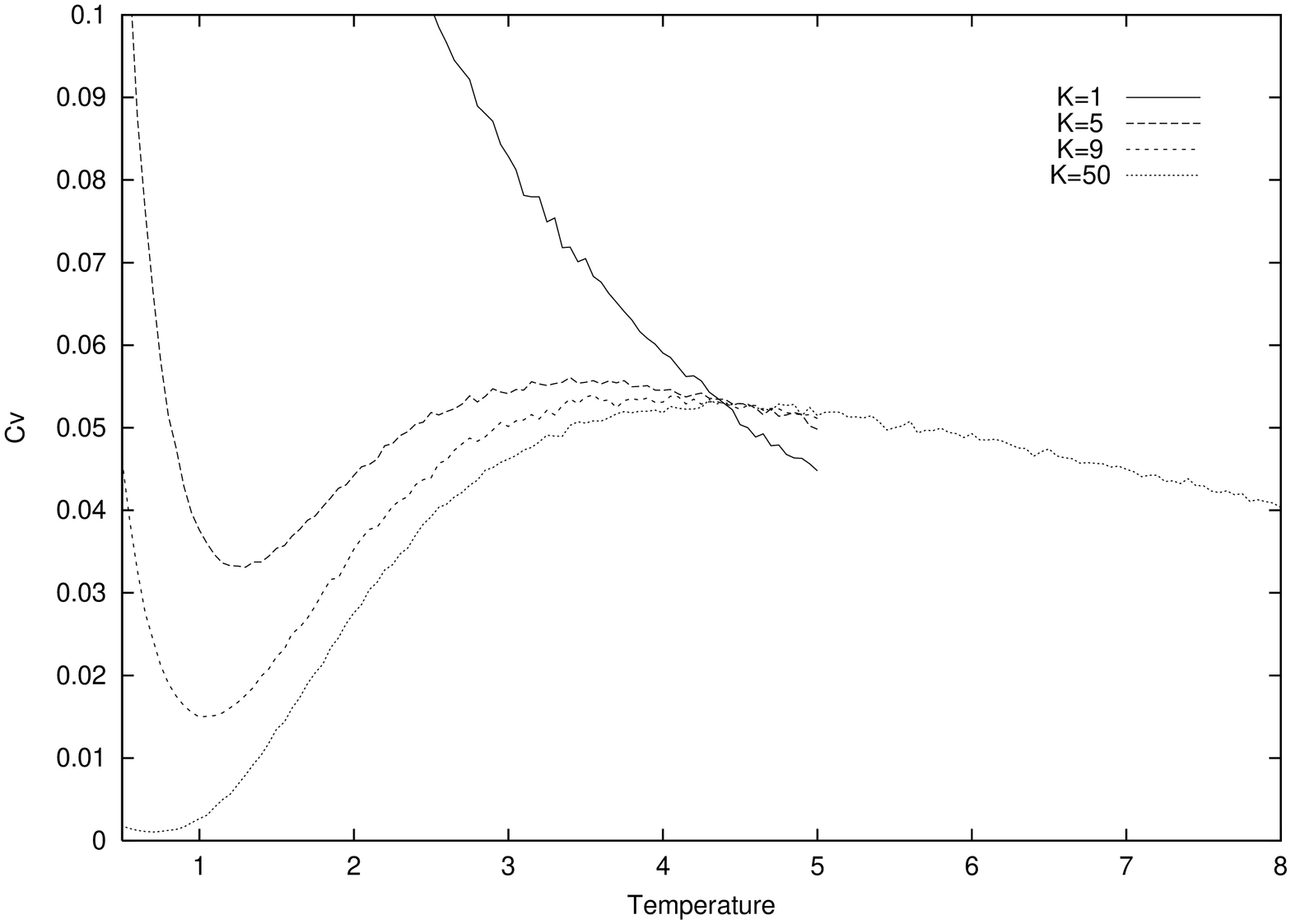}}
\subfigure[]{\includegraphics[scale=0.29,angle=-90]{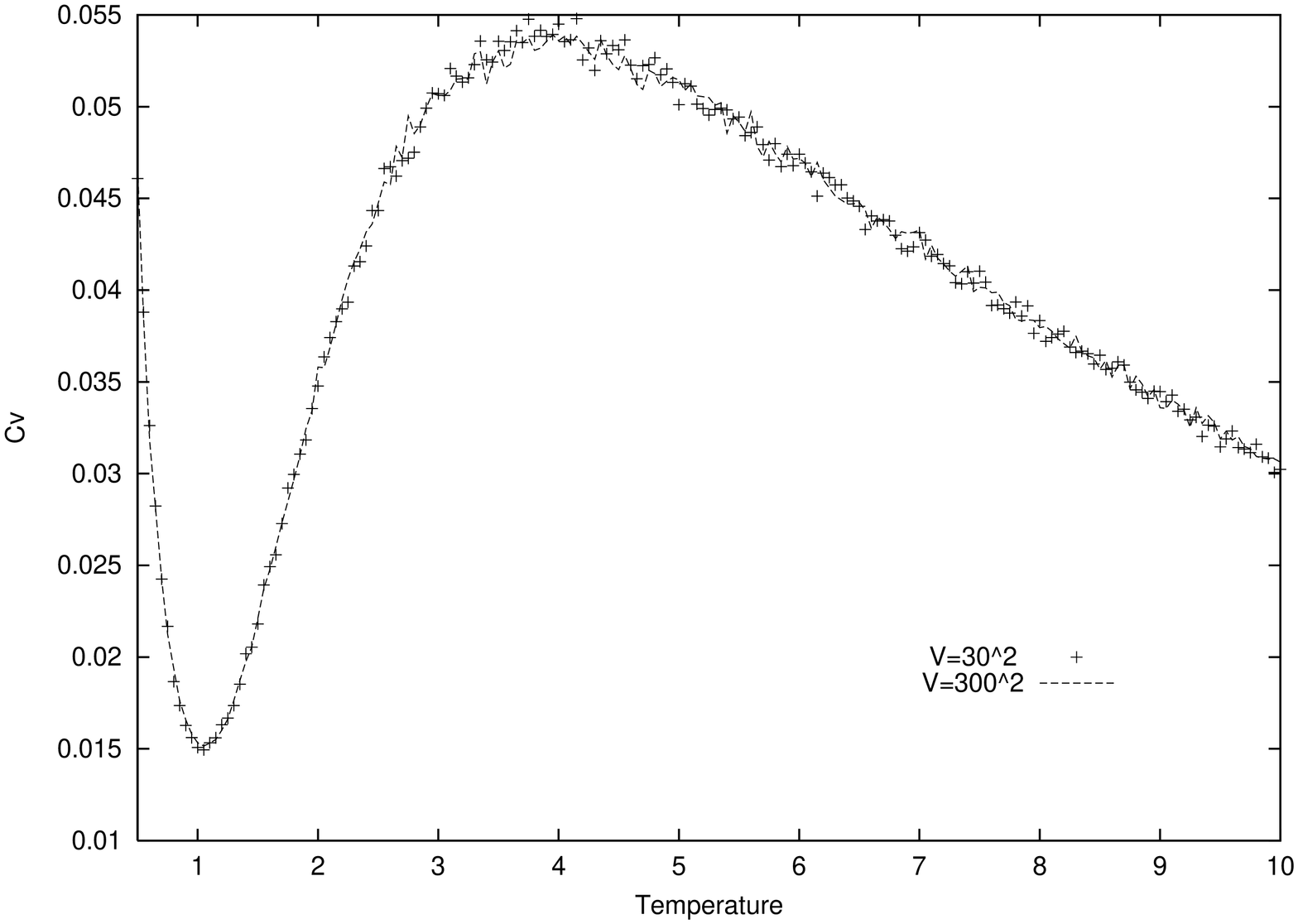}}
\end{center}
\caption{Closer look to the secondary structure. In (a) there is the
  evolution of the structure with $\kappa$. Remember that all the
  data has been rescaled to make the comparison between
  them easier. In (b) the dependence of this structure with the volume is
  tested for $\kappa=9$. The conclusion is that it is not dependent on
  the volume.}   
\label{secondarystructure}
\end{figure}

The same model but in 3 dimensions exhibits a quite
complex phase space, with a thermodynamical transition at a
temperature $T_c$ between two distinct phases that happens to change
from first order to second order when the value of $\kappa$ crosses
some critical value \cite{Ba-es-jo-ma}. Also a dynamical transition is
present in the 3 dimensional model at a temperature $T_g \le T_c$. 

The fact that there is no phase transition in this spin model can be
eventually traced back to the fine tuning of the parameters in the
Hamiltonian. Since there is only one phase, no useful order parameter
can be constructed. This makes impossible the analysis of the dynamics
of this model along the conventional lines of domain growth used in
\cite{Lai-maz}. 
The dynamical properties of this 2D model will be discussed in the next 
section.

\section{Dynamical analysis of the model}\label{dynamics}

As we have mentioned above our motivation in studying this model is to
determine if the dynamical behavior that it exhibits glassy features, as
its 3D  counterpart, or just signs of very slow evolution. 
The technique we shall use in this section will be based on two-time
correlation functions \cite{Al-franz-ri,Bar-bu-me}. Before entering
into details let us stop for a moment to  understand which are the
main features of the evolution of the system. 

\subsection{Thinking about dynamics}

For this qualitative analysis we are going to use the loop language.
As we have seen, all the energy is concentrated in the corners of the
loop and in the crossing of one loop with each other. 
To simplify the reasoning we are going to use the $\kappa=0$ limit
where the loops can freely cross each other, but the same conclusions
can be obtained with $\kappa\neq 0$. 
We are going to study the evolution at low temperatures, so we have to
accept that thermal fluctuations are rare. 

\begin{figure}[!ht]
\begin{center}
\subfigure[]{\includegraphics[scale=0.45]{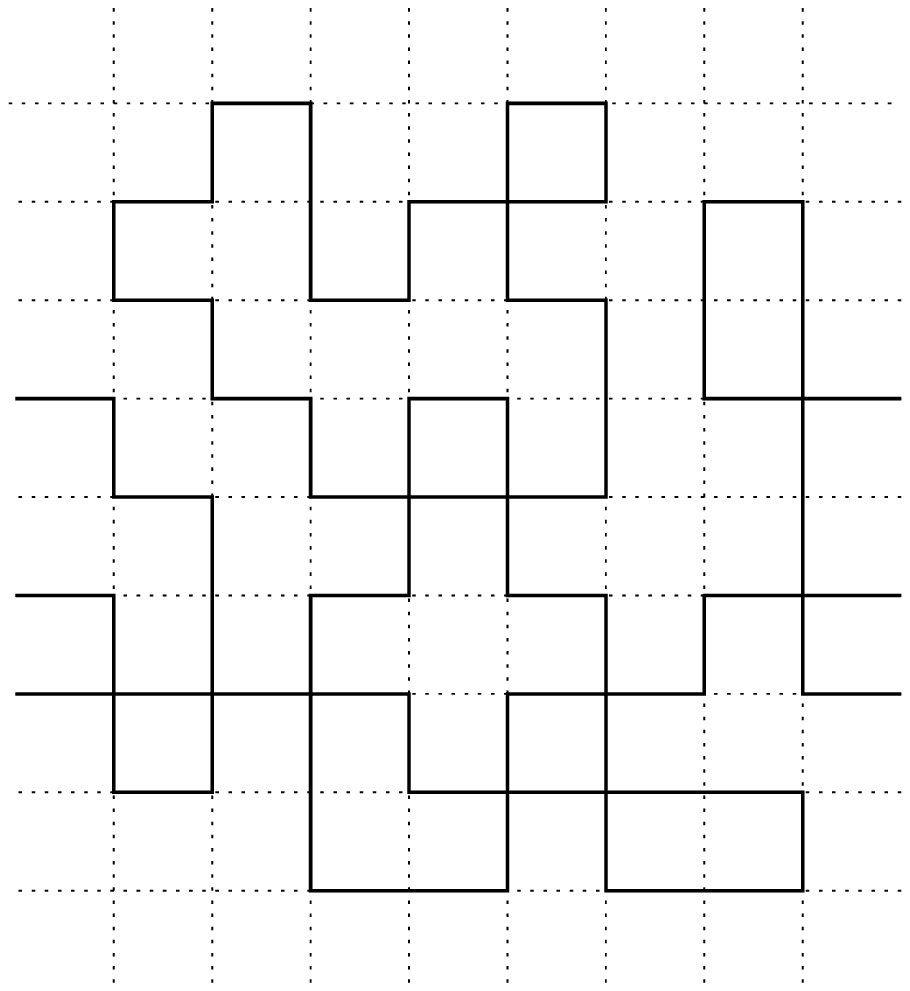}}
\subfigure[]{\includegraphics[scale=0.45]{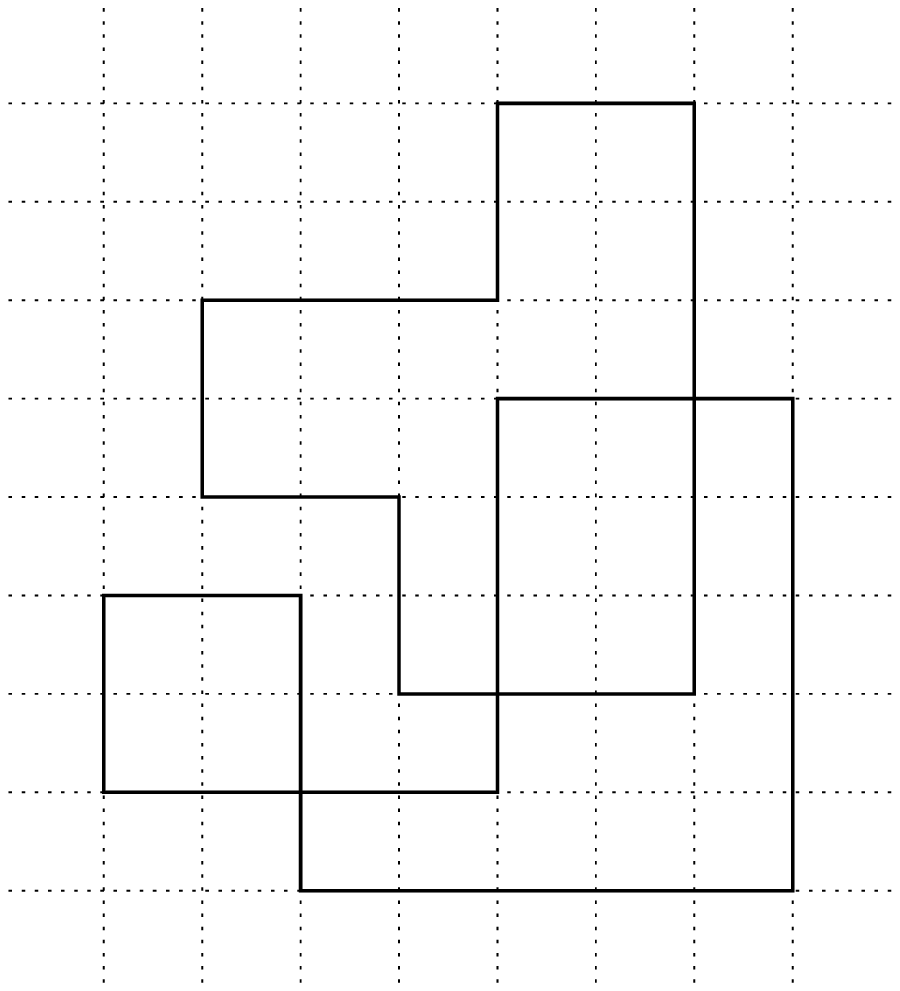}}
\end{center}
\caption{Examples of loop configurations. (a) A disordered one, before
evolution takes place. (b) After some evolution the loops have reduced
the number of corners and have found a metastable configuration.}
\label{orddisord}
\end{figure}

A disordered configuration (fig.\ref{orddisord}(a)) will try to evolve
by straightening the boundaries of the domains
 in order to minimize the number of corners. After this first
thermalization, the system will end up with some long lines glued
together with some corners in a non-optimal 
way (see fig.\ref{orddisord}(b)).   
In general, by decreasing the energy in every step, the loop is
going to get trapped in some very stable states whose
energy  cannot decrease further without increasing it temporarily.
The first phase of the evolution is really fast due to the fact that
almost all moves decrease the energy. 

From this point on, the evolution is quite slow because there are energy
barriers to jump over that the system has created during the first fast
evolution.  

\subsection{Is there a dynamical transition?}

Let us make now a more detailed quantitative analysis of the dynamical
behavior of the model. The magnitude we are going to use is the
autocorrelation function of the energy per plaquette $e_i$
\begin{equation}
C(t,t_w)=\frac{1}{N}\sum_i e_i(t_w)e_i(t)\quad,\qquad t >t_w
\end{equation}
where the sum runs over all the plaquettes in the lattice. To avoid
over-counting  the bonds, we have taken the following definition for
the energy per plaquette. For each plaquette we will count the energy
coming from the plaquette term, the two next to nearest neighbor
terms, and two of the four nearest neighbor terms in such a way that
one bond is horizontal and the other is vertical.  

Let us now describe the results from our numerical analysis. All
simulations shown here have been performed with a metropolis-like
Monte Carlo algorithm with periodic boundary conditions. The volume is
$100^2$ in all the data, unless otherwise indicated. The data
presented in this section correspond to averaging over 25 independent
systems.  

\begin{figure}[!ht]
\begin{center}
\subfigure[]{\includegraphics[scale=0.28,angle=-90]{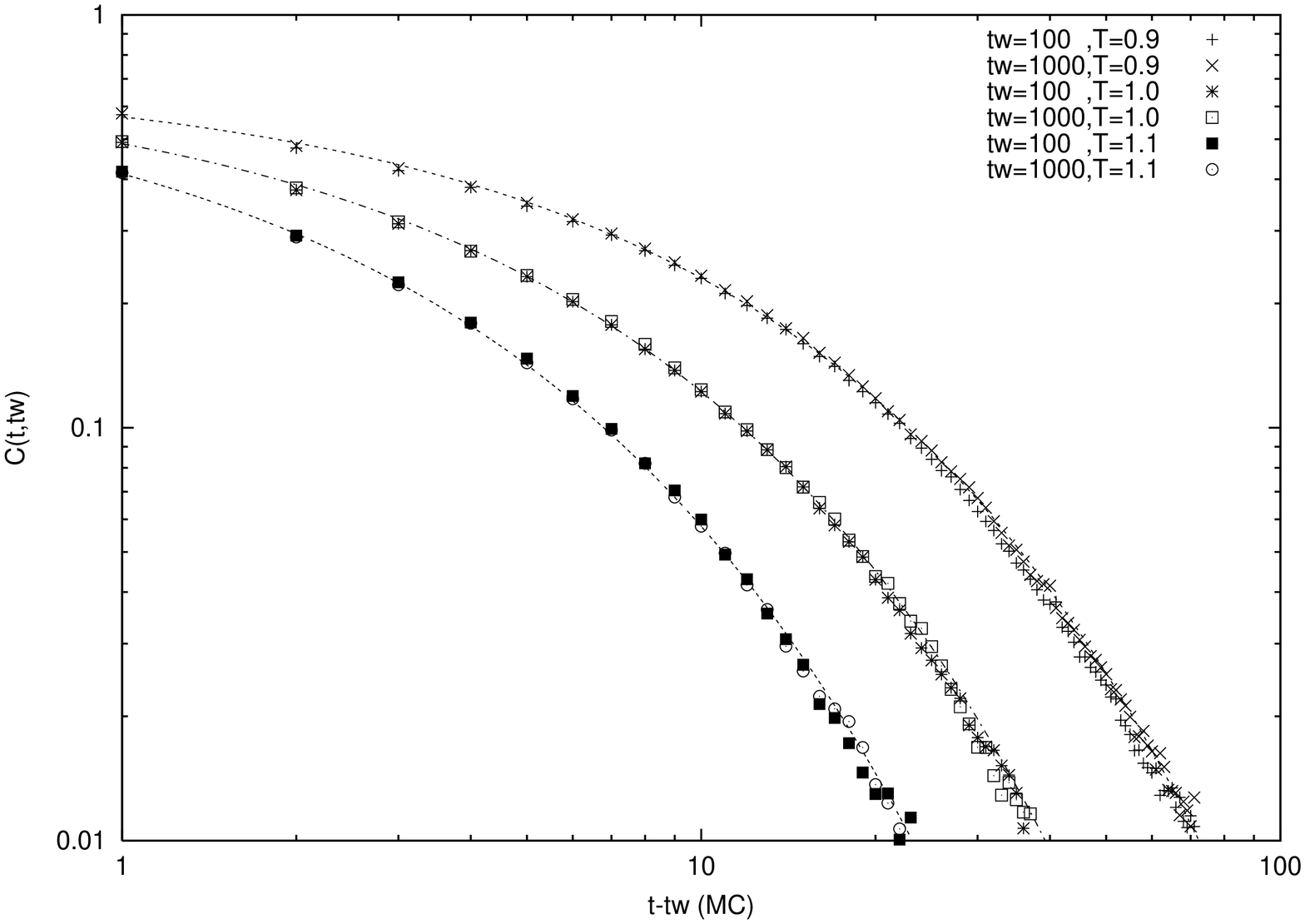}}
\subfigure[]{\includegraphics[scale=0.28,angle=-90]{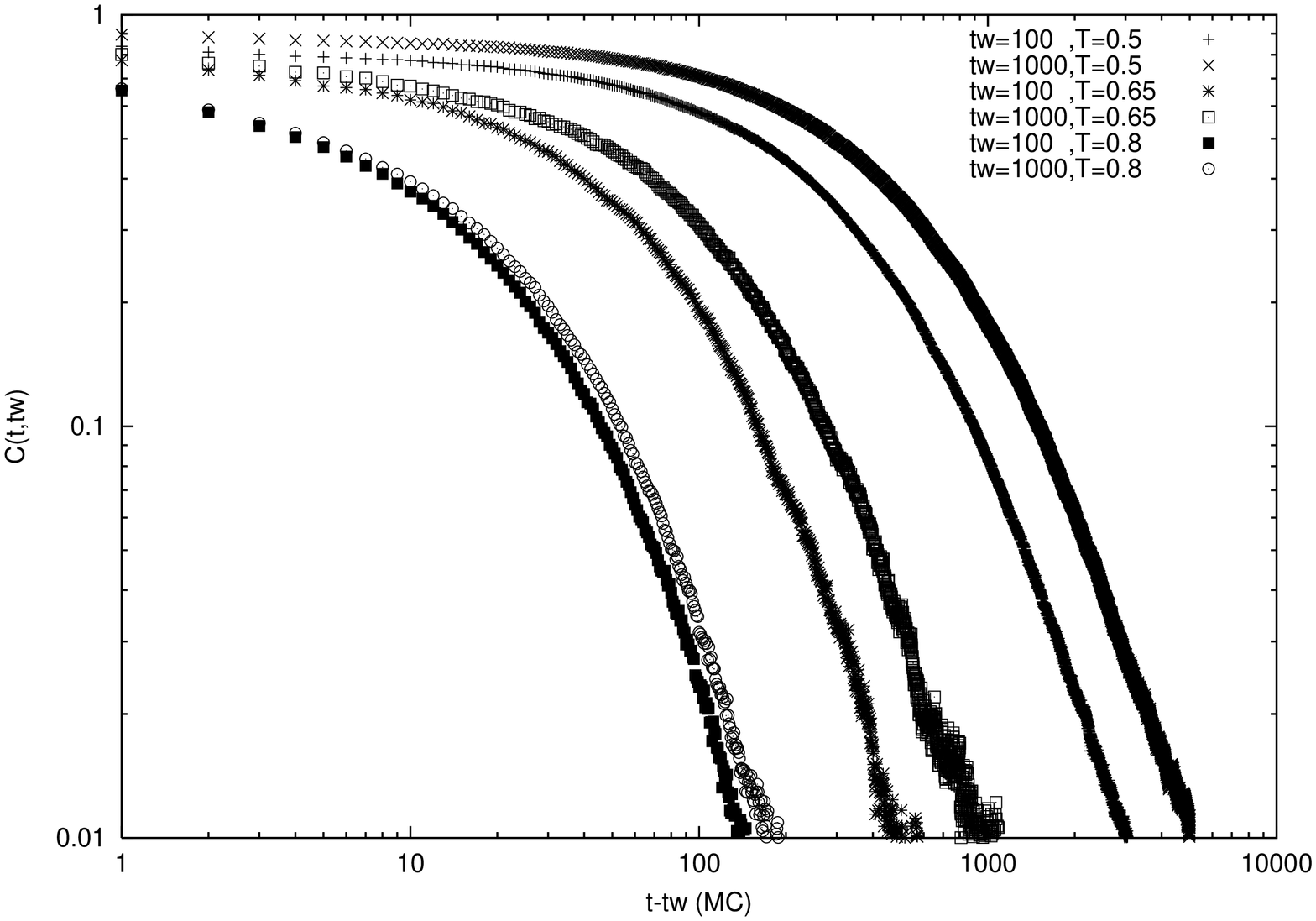}}
\end{center}
\caption{Autocorrelation function for different temperatures. (a) At
  this temperatures the autocorrelation function depends only on
  $t-t_w$. The fits to the data are plotted with lines. (b) At these
  lower temperatures the autocorrelation looks dependent of $t_w$ and
  $t$ independently.} 
\label{correlation1}
\end{figure}

We start by studying two different waiting times like for example
$t_w=100,1000$ and a few temperatures. We can easily see that there
are some temperatures where the autocorrelation function 
$C(t,t_w)$ depends only on $t-t_w$, a good indication that the system
has reached equilibrium (unlike for instance in a glassy phase), while
at lower temperatures the autocorrelation function happens to
depend on $t$ and $t_w$ independently. In figure \ref{correlation1} we
can see an example of this.  
This behavior could hint to the existence of some kind of dynamical
transition like the same model in 3D. To
make it clearer we can look at the form of the autocorrelation
function above $T^*$, where the supposed dynamical transition would
take place. We can attempt a fit to this data with an stretched exponential  
\begin{equation}\label{fitingfunc1}
A\exp\bigg[-\bigg(\frac{t-t_w}{\tau}\bigg)^c\bigg]
\end{equation}
It's clear from the plots in fig.\ref{correlation1}a that the fits are
apparently very satisfactory. 

If we extract $\tau$ from the fits and make a plot as a function of
the temperature, we will see that $\tau$ grows when we decrease the
temperature. This would suggest that the $\tau$ could diverge at some
finite temperature, so we try to fit it with a power-like divergence
function. 
The fitting function we have used is
\begin{equation}\label{fitingfunc2}
\frac{K}{(T-T^*)^b}
\end{equation}
The fit is shown in figure \ref{newtaudT} (solid curve), and it
 provides a value for  $T^*$. The problem is that the value the fit
 delivers is around $0.57$, while looking at fig.(\ref{correlation1})
 we expected something between 0.8 and 0.9, exactly where we are
 beginning to see $t_w$ dependence. The procedure is thus not
 self-consistent and we need some explanation for this discrepancy. 

\begin{figure}[!ht]
\begin{center}
\includegraphics[scale=0.58]{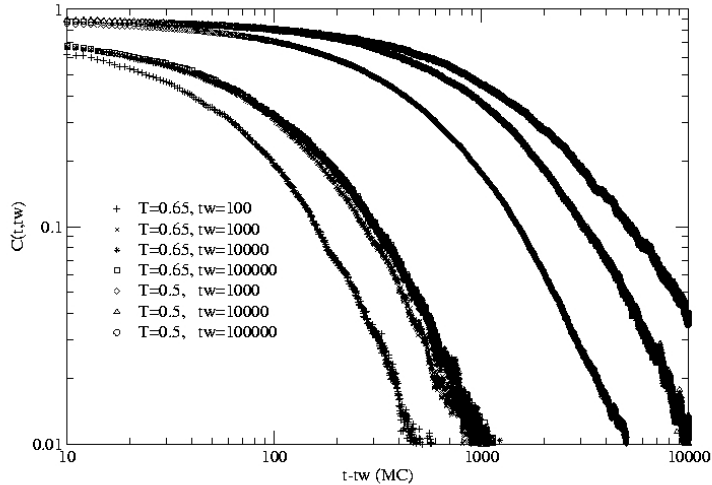}
\end{center}
\caption{Exploring longer waiting times $t_w$ we can see the slow
  approximation to equilibrium of the autocorrelation function. Two
  different temperatures are plotted. In the lower temperature
  ($T=0.5$) the equilibrium is not yet reached, but the range where the
  dependence in $t_w$ is not manifest grows with $t_w$.}
\label{correlation2}
\end{figure}
Let's explore much longer waiting times. If we do that, we will be able to 
understand exactly what is happening. In figure \ref{correlation2} we
discover that at longer waiting times the dependence in $t_w$
disappears, leaving only a function of $t-t_w$. This is an indication
that the system is not in a putative glassy phase but is just
exhibiting an extremely slow relaxation to equilibrium. 

Once we have reached the equilibrium at lower temperatures we can fit
and extract the autocorrelation time. Adding this new data to the
$\tau$ vs. temperature plot we realize that the previous fit is not
satisfactory with this new data, so we are lead to make a new
fit. After this new fit with more data is performed, the new value for
$T^*$ happens to be much lower than the previous estimation (see
fig. \ref{newtaudT}, doted curve).The new value of $T^*$ 
\begin{figure}[!ht]
\begin{center}
\includegraphics[scale=0.4,angle=-90]{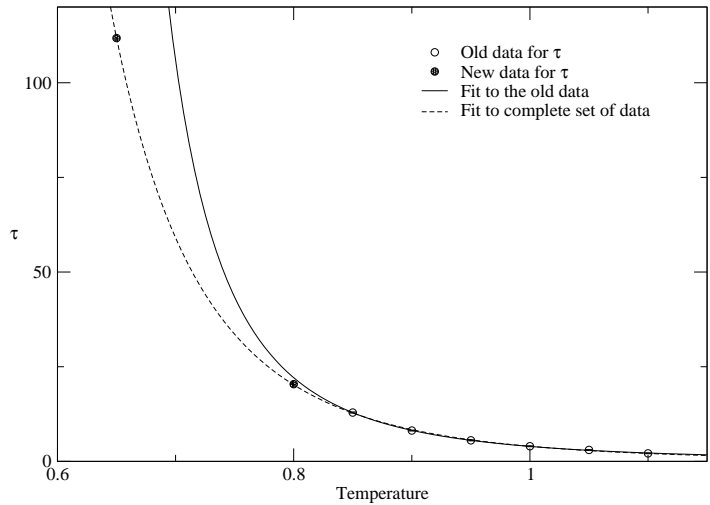}
\end{center}
\caption{New data and new fit of the autocorrelation time in terms of
  temperature.} 
\label{newtaudT}
\end{figure}
decreases to $0.29$. Thus supposing that we can go on equilibrating
the system at any temperature for large but finite values of $t_w$, we
must conclude that the $T^*$ parameter will get closer to zero with
each new point we include in the data.  
We conclude that there is no dynamical transition to a glassy phase,
even though that was the case in the 3 dimensional version of the model.

\begin{figure}[!ht]
\begin{center}
\includegraphics[scale=0.58]{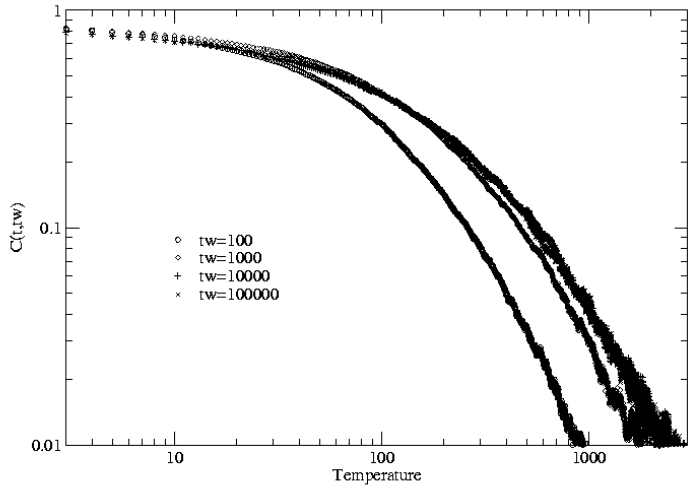}
\end{center}
\caption{Example for $\kappa=9$ of the approximation to the
  equilibrium of the autocorrelation function at temperature $T=0.7$} 
\label{correlation3}
\end{figure}
The autocorrelation function at low temperatures depends on $t_w$
but when we increase the value of $t_w$ this dependence disappears
completely. In this example the two last sets of data for $T=0.65$
coincide, so we can declare that it is independent of $t_w$ for $t_w >
10^4$ (at this temperature). We have not reached the complete
equilibrium in the $T=0.5$ case, but we can see that for $t_w=10^4$
and $t_w=10^5$ the coincidence has grown considerably. So the
conclusion is that the autocorrelation is approximating to some
equilibrium shape.
For $\kappa\neq 0$ the analysis follows exactly the same steps,
and the same conclusion is reached. 
We can see in fig.\ref{correlation3} that the
same kind of behavior is present in $\kappa=9$. 
 
We can take a look at other observables like the two time overlap
function $Q(t_w+t,t_w+t^{\prime})$, or the autocorrelation of the
local magnetization $C_m(t_w+t,t_w)$ \cite{Bar-bu-me}. Suppose we let a system evolve
through a time $t_w$, then we make two copies of the system and evolve
them independently $t$ and $t^{\prime}$ respectively, then the
observables are defined as
\begin{eqnarray}
Q(t_w+t,t_w+t^{\prime})&=&\frac{1}{N}\sum_i \sigma^{(1)}_i(t_w+t)
\sigma^{(2)}_i(t_w+t^{\prime}) \\
\qquad C_m(t_w+t,t_w)&=&\frac{1}{N}\sum_i \sigma_i(t_w)
\sigma_i(t_w+t)
\end{eqnarray}
where the upper-index indicates which is the copy that the spin belongs
to.

\begin{figure}[!ht]
\begin{center}
\includegraphics[scale=0.58]{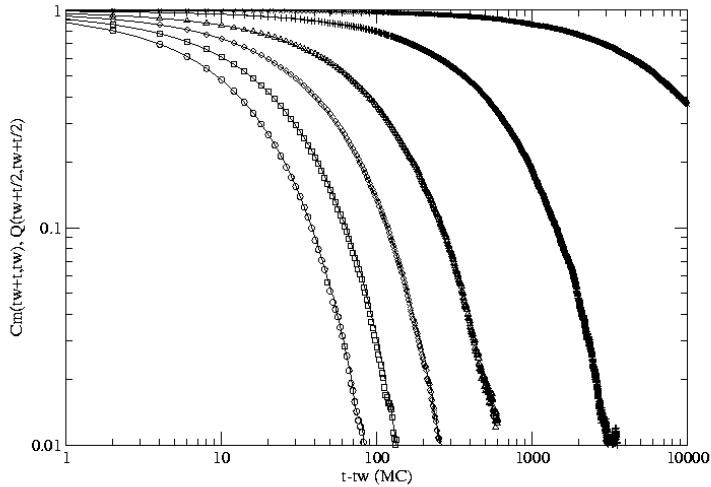}
\end{center}
\caption{The autocorrelation function of the local magnetization
  $C_m(t_w+t,t_w)$ (Symbols) and the overlap function
  $Q(t_w+t/2,t_w+t/2)$ (lines). Lines are $C_m$ and symbols are $Q$
  data. From left to right, $T=1.1$($\bigcirc$), $T=1.0$($\square$),
  $T=0.9$($\Diamond$), $T=0.8$($\triangle$), $T=0.65$(+),
  $T=0.5$($\times$). Formula (\ref{replica}) is clearly verified.} 
\label{CandQ}
\end{figure}

In equilibrium (that is when the autocorrelation is independent of
$t_w$) they should satisfy 
\begin{equation}
\label{replica}
Q(t_w+t,t_w+t)=C_m(t_w+2t,t_w).
\end{equation}

We can see in figure \ref{CandQ} that the relation (\ref{replica}) is fully accomplished by our system. Another indication of the
non-glassiness of our model (for $\kappa=0$ in this case). The same
behavior is present in $\kappa \neq 0$. 

\section{Analytical results for the evolution}\label{longtimes}

One of the differences between glassy and non-glassy evolution
is the fact that for the former, logarithmic growth of domains
dominates the evolution of the system at long times. Then  we are used
to talk in terms of domains and domain walls, velocity of the domain
growth, or the energy contained in a domain wall.  

The reason that the domain growth concepts cannot be applied to the
gonihedric model in 2 dimensions, unlike traditional Ising-type models,
is that there 
is no good local order parameter
that allows us to say when a piece of `ordered' system is in one
ground-state or the other, so we cannot distinguish domains with
different ground-state configuration in its bulk. 

\begin{figure}[!ht]
\begin{center}
\subfigure[]{\includegraphics[scale=0.6]{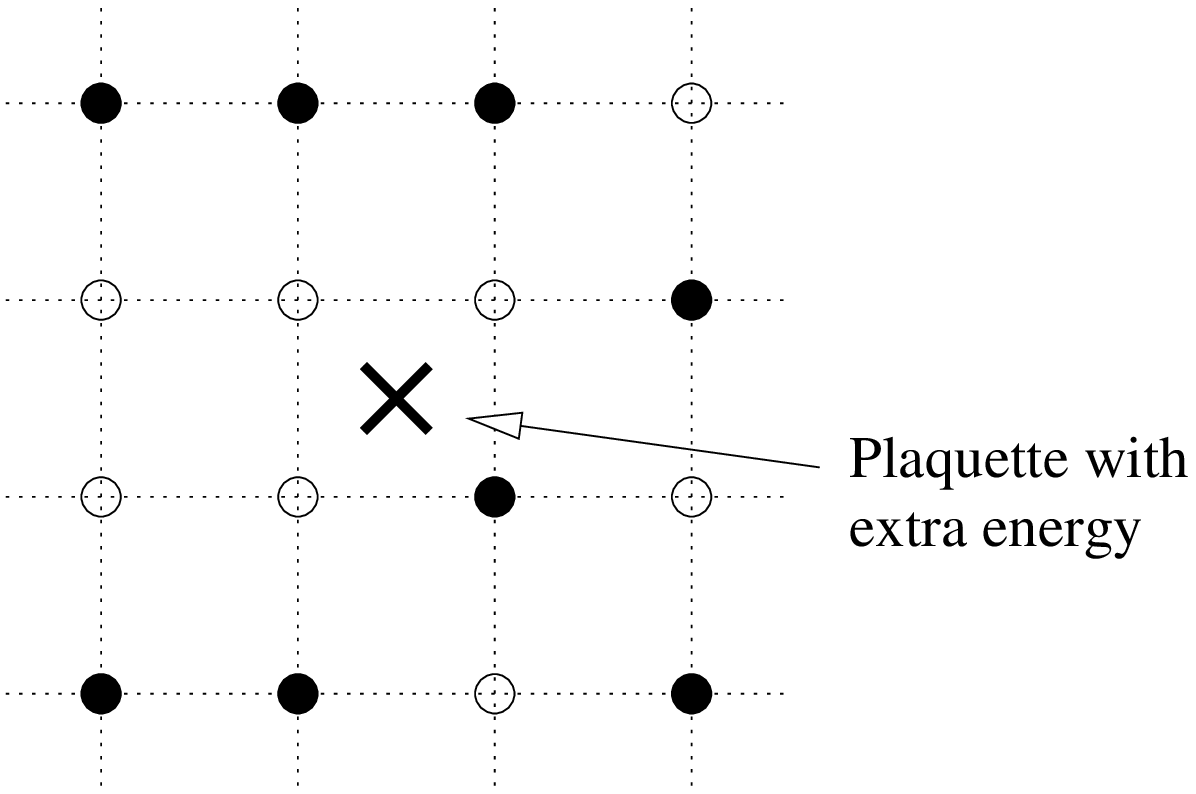}}
\subfigure[]{\includegraphics[scale=0.6]{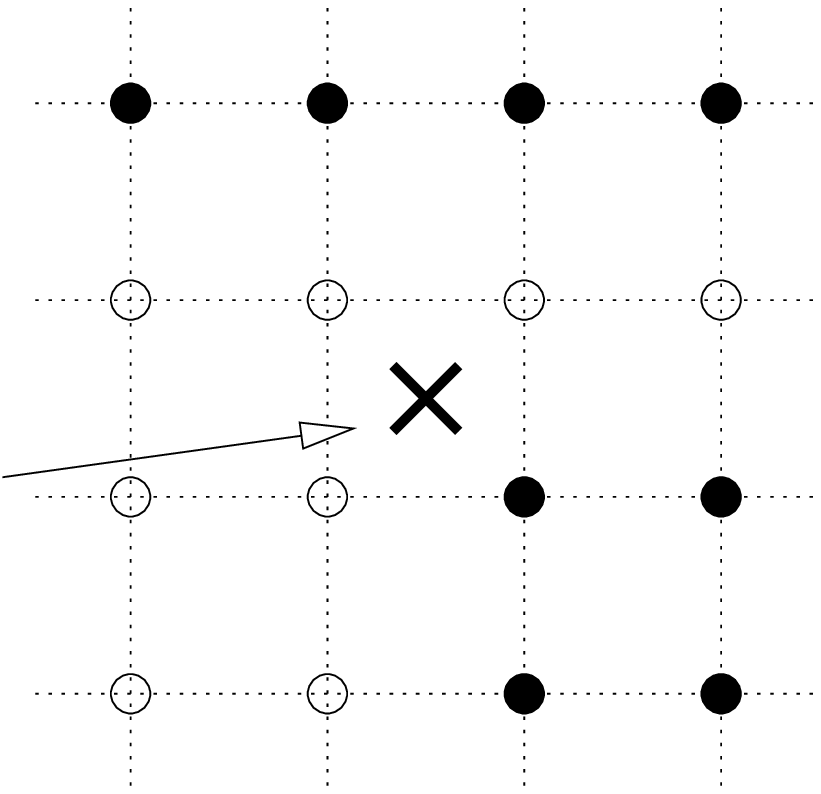}}
\end{center}
\caption{Examples of isolated defects for (a) $\kappa=0$ and (b)
  $\kappa\neq 0$}  
\label{nodomainex}
\end{figure}

In the gonihedric spin model there are so many different
ground states that we can travel around a plaquette without crossing
any extra accumulation of energy, and yet find extra energy in the
plaquette we have been surrounding. This would not be possible if a
domain wall would have existed. Here rather than domains we have to talk 
about point-like defects. In figure \ref{nodomainex} we can see an
example for $\kappa=0$ and for $\kappa\neq0$ where an isolated defect
(accumulation of energy) is marked with a big $\times$.  

\subsection{Defect dynamics at very long times}

From now we are going to consider the case $\kappa=0$.
Already in \cite{Bu-gar} Buhot and Garrahan defined the dual version
of the gonihedric model we are going to use in this section. This
duality is just a change of variables, from spins to plaquette
variables, where the plaquette variable can be defined in the way 
\begin{equation}
\tau_{[i,j,k,l]}=\frac{1-(\sigma_i\sigma_j\sigma_k\sigma_l)}{2}
\end{equation}
\textit{i.e.}, the plaquette dual variable is zero when there is no
extra energy accumulated on it and is equal to $+1$ when there is a
defect there. Then the extra energy of the system will be just the sum
of the $\tau$ variables, or the number of defects\footnote{Note that
  this description is useful at long times and low temperatures only
  when $\kappa=0$. This is the cases where the
  crossing-loop-like plaquettes do not contribute to the energy.}.

But the dual model is not just a model of defects. The fact that the
defects are defined in terms of spin configurations of an interacting
spin model is essential, and provides the defect model an special
constrained dynamics. 
There are some rules in order to move, create or annihilate defects: 
One defect on its own can't move, it is stable as it is, the only way
it can move is through the creation of two more defects, that means to
climb up an energy barrier $\Delta E=2$. In contrast, two neighboring
defects can freely move, but only in one direction (horizontal pairs
move vertically and vertical pairs move horizontally). 
The only way defects can disappear is by meeting four defects in a
square pattern, or when a moving pair collides with an isolated defect,
then the moving pair will disappear moving the isolated defect as a
result. This description in terms of moving defects will allow us to
find an analytical expression for the energy decay at very long times.

The energy is related with the number of defects as we mentioned
before, so we would like to know how defect density evolve in time. To 
do that we need to understand which is the leading mechanism that
make defects disappear.

Our starting point will be a system that has relaxed from a disordered
configuration to a low temperature for a long time. So at that moment
we have to consider that all the defects are isolated. In these
conditions, the movement of all those defects is really slow, because to
move they have to create a pair of defects, \textit{i.e.} overcome an
energy barrier. The probability to do this is $\sim \exp(-2/T)$, where
$T$ is the temperature. The characteristic time will this be $\sim \exp(2/T)$. 
For low temperatures, this will be very long time, and we can neglect the possibility 
of two pairs of defects being created successively. 

\begin{figure}[!ht]
\begin{center}
\includegraphics[scale=0.6]{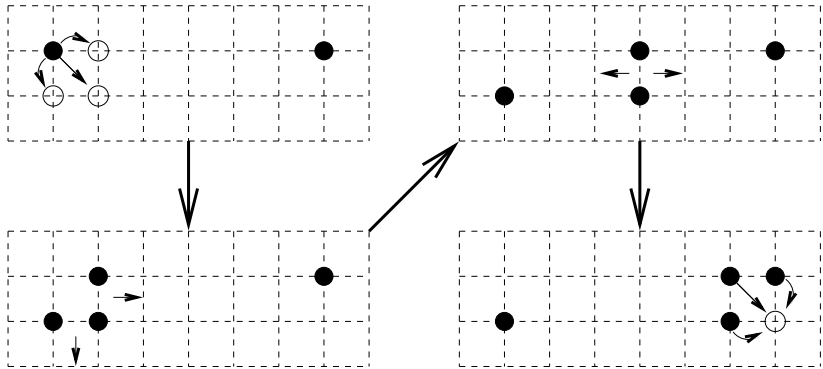}
\end{center}
\caption{Sketch of the leading process that allows defects to
  move. In the first step an isolated defect transforms in three
  complementary defects, then two of them move in a random walk like
  way until they find a second defect to combine and disappear.}   
\label{randomstep}
\end{figure}

Let us assume that a pair of defects has been created (see the first 
diagram of Fig. \ref{randomstep}).
After this creation, two defects will move freely in horizontal or
vertical direction. Because the move of the pair is much faster than
the creation of the next pair, the process may conclude in two ways: either the
pair returns to the defect it just left behind and returns to the 
original configuration, or it finds in its 
random-walk-like movement another defect, collides with it, and
disappears, resulting in a move for the secondary defect too. The first
case leaves the final configuration unchanged so it represents a
frustrated trial of moving an isolated defect, while the second 
ends up with two defects displaced by one lattice step. In figure
\ref{randomstep} there is an sketch of that process.  

Thus as we are not sure that creating that extra pair is going to
provide a move of the defect because of the frustrated trials, we cannot 
compute the average time for the traveling pairs of defects to reach a given 
distance $d$. In fact this average is divergent\footnote{This can be easily seen by setting the starting point at $x_0=1$, the absorving wall at $x_0=0$, and the target at $x=2$. 
The average time to reach $x=2$ will be $\langle T\rangle=\infty\times\frac{1}{2}
+1\times\frac{1}{2}$}.
We have to compute instead the probability of a success in colliding with a different
defect once a pair is created. The inverse of this probability will give to us (by the same argument as before) the characteristic time for the successful move to happen.

The probability of creating a pair is already known and is
$\sim\exp(-2/T)$, the probability of a successful move of the pair {\it i.e.} reaching another defect, can be easily determined by considering a random walk with an
absorbing wall at the origin, and computing the probability of arriving at a distance $d$ for the
first time \cite{Fish}. Some details are given in Appendix \ref{randomwalkwall}. 
The result is $1/d$. Thus the final probability for an isolated
defect to move one step is
\begin{equation}
\mathcal{P}\sim \frac{\exp(-2/T)}{d}
\end{equation}
But the distance $d$ can be parameterized in terms of the density of
defects $\rho$ like $d\sim 1/\rho$ so
\begin{equation}
\mathcal{P}\sim \rho\exp(-2/T)
\end{equation}
and clearly the number of MC steps needed to move one site a given
defect is
\begin{equation}
\tau_1\sim\mathcal{P}^{-1}\sim \frac{\exp(2/T)}{\rho}
\end{equation}

Now, we know the probability of an isolated defect to move one step in the 
lattice. The next thing we need to know is how often two defects meet each other and become a pair.
This is interesting because once they become a pair they will 
move freely and will easily find a third defect to decay with. Only at
this point and not before we have decreased in two units the number of 
defects. Let's do the calculation. 

As the move of a pair is very fast, we only need to know the time
needed for one defect to meet another one. The move of defects is a 
slow random walk with a characteristic time click $\tau_1$. Unlike in the 
previous case the probability of one defect meeting another one is $1$, and the characteristic time needed to travel a distance $d$  will be proportional to $d^2$, so finally the characteristic time $\tau$ needed for two defects to meet is 
\begin{equation}
\tau\sim\tau_1\bigg(\frac{1}{\rho}\bigg)^2\sim\frac{\exp(2/T)}{\rho^3}
\end{equation}

Now we can set the differential equation of the evolution of the number
of defects 
\begin{equation}
\label{diffeq}
\frac{\textrm{d}\rho}{\textrm{d}t}\sim\frac{-2}{\tau}\rho
\sim-2\rho^4\exp(-2/T) 
\end{equation}

This differential equation is valid only for low
temperatures and long times (because there are only isolated defects),
and low density of defects (because we considered large distances
between defects), but this condition is implicit if we demand low
temperatures and very long times.

Solving (\ref{diffeq}) we find that the density of defects should
evolve in time like $\rho\sim t^{-\frac{1}{3}}$ and as a consequence
the energy evolves in the same way. In the next section we are going
to perform a simulation of the energy at very long times and compare
the way it evolves with our prediction. 

\subsection{Long times simulation}

To compare with the analytical result we performed long simulations
at very low temperatures. For this purpose we started with a
disordered initial configuration and let it evolve with a Monte Carlo
algorithm at very low temperatures like $0.4$ or $0.33$. The final
data is the average of 20 independent evolutions from 20 different
initial states.

\begin{figure}[!ht]
\begin{center}
\includegraphics[scale=0.58]{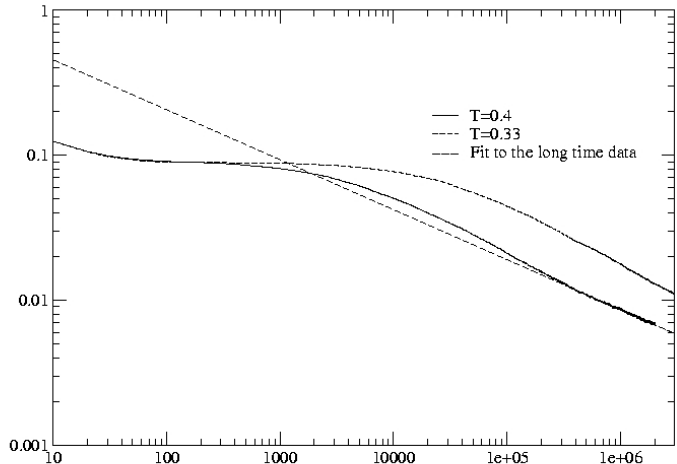}
\end{center}
\caption{Defect concentration in function of time in a log-log
  plot. The slope at the latest stage of evolutions around 0.34 for
  $T=0.4$.}   
\label{defectes}
\end{figure}

In figure \ref{defectes} we can see the evolution in time of the
defect concentration\footnote{This plot is in complete agreement with
  the plot in fig. 2a of ref.\cite{Bu-gar}, where different aspects the
  same model are analyzed with a different kind of Monte-Carlo
  algorithm. Note that our temperature scale is related to the one
  in \cite{Bu-gar} by a factor of $2$ coming from the Hamiltonian we
  used in our simulations.}, closely related to the energy density
through the relation $\rho=(E+1)/2$ where the energy density is
defined here as 
$E=-\frac{1}{N}\sum_i(\sigma\sigma\sigma\sigma)_{\Box_i}$ 

The plateau starts when \cite{Bu-gar} the system has already reached
an stable configuration and finds energy barriers that makes difficult
to decrease the energy. As we have seen before those energy barriers
cost an energy $\Delta E=2$, so the time needed to reactivate the
evolution will be of order $\sim \exp(2/T)$. After the plateau the
evolution contains only isolated defects and spontaneous fluctuations
in form of pairs of defects that appear when an isolated defect is
trying to move. So this should be the range of validity of our
calculation, or in other words, in this region the evolution should be
like $\rho\sim t^{-\frac{1}{3}}$. Note that this behavior should set
in rather slowly (see Appendix\ref{randomwalkwall}) and therefore it
should be apparent only for very low concentration of defects. Note
also that the evolution depends only on the concentration of defects
once we are in the activated regime.

Indeed when we look at plot of the data (see fig.\ref{defectes}) it is
clear that in the activated regime in a logarithmic scale, the
behavior is approximately linear. However the slope changes slightly
with the concentration of defects, which we understand as a signal of
the slow setting in of the asymptotic regime which we just
discussed. At this very late stage (beyond $\sim 5\times 10^5$
Monte-Carlo steps for $T=0.4$) the slope stabilizes to a value close
to $-0.34$; that is really close to the one we predicted. Note that
the bulk of the data lies precisely in this region (for this
temperature we have run up to $2\times 10^6$). 
For $T=0.33$ we have not yet reached the region where the fit of the
slope becomes stable, in spite of having run up to $3\times 10^6$
sweeps; however we have compared the slopes at similar
values of the concentration of defects with the $T=0.4$ case and found
quite similar values. From this we conclude that for enough long times
we would obtain a value for the slope compatible with the $-0.33$ we expect. 

\begin{figure}[!ht]
\begin{center}
\includegraphics[angle=-90,scale=0.4]{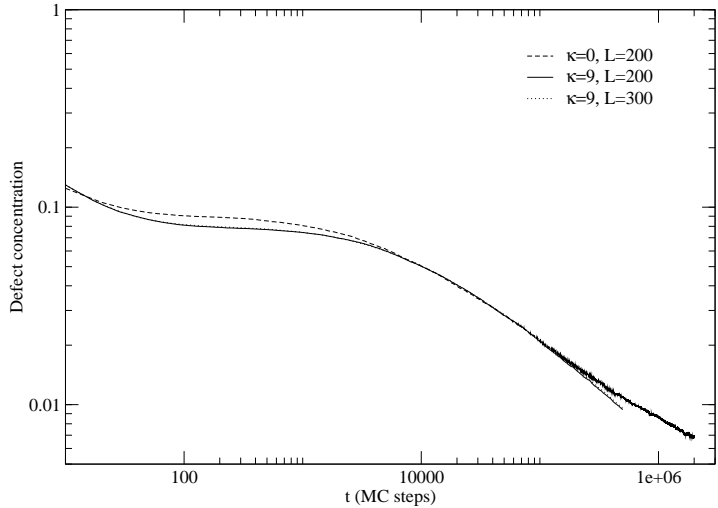}
\end{center}
\caption{Defect concentration in function of time in a log-log
  plot for $T=0.4$. Two different volumes for $\kappa=9$
  and one for $\kappa=0$ are compared.} 
\label{defectsDK}
\end{figure}

For $\kappa\neq 0$ it is harder to know exactly what is the
law for the evolution of the defects at long times and low
temperatures. Some simulations have been made. In figure
\ref{defectsDK} we compare a long time simulation of $\kappa=0$ and
$\kappa=9$. It can be seen that the $\kappa=9$ case does not seem to
follows a power law (two different volumes for $\kappa=9$ are shown to
reflect that the plot is volume independent). At very long times the
evolution, though rather similar, it is actually faster, since the
defect concentration is reaching lower values in shorter times. This
is expected due to the lack of a pure geometrical interpretation in
the $\kappa\neq 0$. 

\section{Conclusions}
\label{conclusions}
In this work we have analyzed the dynamical behavior of a two
dimensional spin model with  very `geometric' couplings. The
microscopic surface tension is zero and the energy is concentrated on
the corners of the loops separating regions of different ferromagnetic
states (ferromagnetic states are not the only ground states, the
degeneracy of the ground state is $2^{2 \ell}$ or $2^\ell$ depending on
whether the self-avoidance parameter $\kappa$ is turned on or not). 

The model has rather trivial thermodynamic properties for non
self-avoiding loops. It can actually be mapped to an exactly solvable
six-vertex model, albeit exhibiting rather remarkable finite size
effects. When the self-avoidance parameter is turned on, no phase
transition or thermodynamic singularity is found.

On the contrary, the dynamical properties of the system are rather
interesting. Its 3D counterpart does exhibit logarithmic growth of
domains and quite clear glassy behavior below a certain temperature.
We do find slow dynamics, but they rather correspond to a power law
($\kappa=0$) or faster ($\kappa\neq 0$), and definitely there is no 
sign of glassy dynamics ---at least down to the rather low
temperatures we have explored. 

\section*{Acknowledgments}
D.E. and A.P acknowledges support from ''EUROGRID- Discrete random
geometries: from solid state physics to quantum gravity''
(HPRN-CT-1999-00161). D.E. acknowledges also support from
FPA-2001-3598 and CIRIT grant 2001SGR-00065, and A.P. from CIRIT grant
2001FI-00387. 

\appendix
\section{The 2D finite size partition function for $\kappa=0$}
\label{finitevolume}

The partition function for our model with $\kappa=0$ is,
\begin{equation}\label{K0ham}
\mathcal{Z}=\sum_{\{\sigma\}}e^{-\beta\ham_{\kappa=0}}
\end{equation}
where $\{\sigma\}$ is the set of all possible configurations of spins,
and $\ham_{\kappa=0}$ is the $\kappa=0$ rescaled Hamiltonian 
\begin{equation}
\ham=-\sum_{[i,j,k,l]}\sigma_i\sigma_j\sigma_k\sigma_l=
-\sum_\Box\sigma\sigma\sigma\sigma 
\end{equation}
the last step just being a simpler form of writing the Hamiltonian,
with the notation $\Box$ meaning spins forming a plaquette in the
lattice. 

We can transform the expression (\ref{K0ham}) in the following way
\begin{eqnarray}
&&\mathcal{Z}=\sum_{\{\sigma\}}e^{-\beta\ham}=\sum_{\{\sigma\}}\prod_\Box
e^{\beta\sigma\sigma\sigma\sigma}\\
&&\quad=(\cosh(\beta))^{ N}\sum_{\{\sigma\}}\prod_\Box\Big\{1+
x\,\sigma\sigma\sigma\sigma \Big\}
\end{eqnarray}
where $x\equiv \tanh(\beta)$
Expanding the product and performing the summation over configurations
only terms with an even power on each spin will survive.
It's not difficult to see that this summation can be mapped into another
combinatorial problem.
\begin{figure}[!ht]
\begin{center}
\includegraphics[scale=0.6]{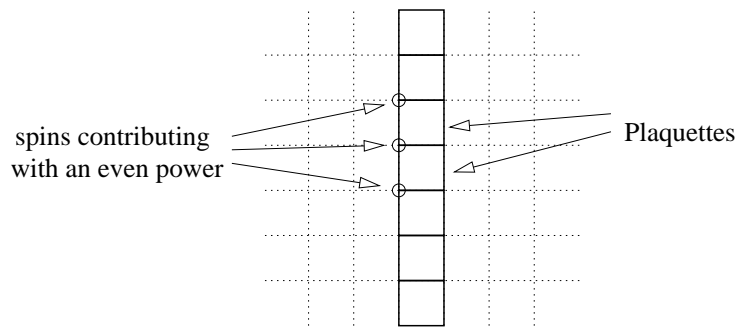}
\end{center}
\caption{one column of plaquettes. Each spin appears 2 times so this
  term contributes to the sum with a weight
  $x^\ell$.}
\label{combinatorial1}
\end{figure}

Consider that we have a term that contains one plaquette. This term will
not contribute unless some of the plaquettes beside it appear also in
that term. We have two ways to make this term contribute, either we
take also the plaquettes above and below it or the plaquettes at the
right side and the left side. In any case we still have problems with
four spins (two spins of each new plaquette), so if we continue adding
plaquettes in the same direction we will complete a vertical row of
plaquettes or an horizontal one (see fig.\ref{combinatorial1}) with
the help of the boundary conditions. That means that the simplest
combination of plaquettes that is going to contribute to the summation
will be a column or an horizontal line of plaquettes, and its weight
will be $x^{\ell}$ where $\ell$ is the length of the row ($N=\ell^2$)  

\begin{figure}[!ht]
\begin{center}
\includegraphics[scale=0.6]{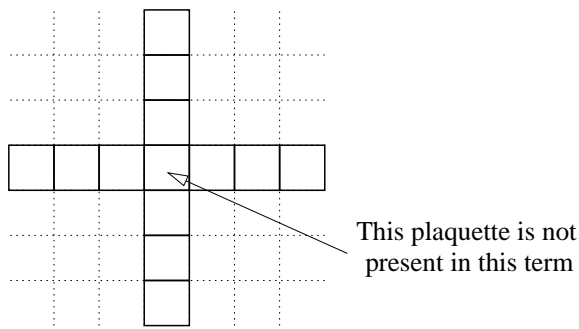}
\end{center}
\caption{An example of a term with vertical and horizontal lines. This
  term contributes with a weight $x^{2\ell-2}$.} 
\label{combinatorial2}
\end{figure}

Then we have to count all possible combinations of vertical and horizontal
lines, multiplying their weights. Only two more things have to be taken
into account. When a term contains horizontal and vertical lines some
plaquettes have to be removed because if not their spins would have an
odd power (see fig.\ref{combinatorial2}). The plaquettes that we have
to remove are the ones on the 
crossings of vertical and horizontal lines. Finally an overall $1/2$
factor has to be used to compensate the over-counting, because each
spin configuration has two possible representations in this
combinatorial problem.

All this has been made to transform the expression 
\begin{equation}
\sum_{\{\sigma\}}\prod_\Box
\Big\{1+\sigma\sigma\sigma\sigma x \Big\}
\end{equation}  
in
\begin{equation}
2^{ N}\frac{1}{2}\sum_{v=0}^{\ell}\sum_{h=0}^{\ell}
\binom{\ell}{v}\binom{\ell}{h}
x^{\ell(h+v)-2hv}  
\end{equation}
that is much more simple at least in the computational sense.
The combinatorial factor $\binom{\ell}{v}$ is the number of
different configurations for $v$ vertical columns of plaquettes, and
the same for the horizontal. Working a little bit more we can simplify 
this expression one step further performing one of the summations and
find its final form  
\begin{equation}\label{lastcombinat}
\mathcal{Z}_{\ell}=(2\cosh(\beta))^{ N}
\frac{1}{2}\sum_{v=0}^{\ell}
\binom{\ell}{v}\Big[x^v+
  x^{\ell-v}\Big]^{\ell}  
\end{equation}
where is now clear the over-counting if you realize that each term 
is invariant under $v\to\ell-v$.

Now this expression can be calculated easily at any temperature with a
computer for any square lattice. Also in the limit
$\ell\to\infty$ we can calculate exactly the sum in
(\ref{lastcombinat}) (which is equal $2$ for any temperature different
from zero) and recover the exact expression for the infinite volume
partition function 
\begin{equation}
\mathcal{Z}_\infty=(2\cosh(\beta))^{ N}
\end{equation}

Now from this expressions for the partition functions we can extract
information like the energy or the specific heat that we plot in
section~\ref{thermodynamics}.

\section{Probability of a pair of defects meeting a third one}  
\label{randomwalkwall}

The magnitude we want to calculate is the probability for a pair of defects following a random walk with an absorving wall at $x=0$ to reach a distance $x$, where the pair is absorbed, starting at point $x_0$. We call this probability $f_n$. In the asymptotic limit where the number of random walk steps $n$ is large, the probability of traveling from $x_0$ to $x$ in $n$ steps is
\begin{equation}
Q_n^0(x,x_0)=\frac{e^{-(x-x_0)^2/2n}}{(2\pi n)^{1/2}}.
\end{equation}
The index zero denotes that this is an unrestricted random walk.
Now we need to find the probability of going from $x_0$ to $x$ in $n$
steps with an absorbing wall at the origin. We shall use the method 
of images in order to subtract the random walks that are forbidden
because of the wall with an auxiliary walker that starts his walk at
position $-x_0$. The probability we are interested in is
\begin{equation}
Q_n^W(x,x_0)=Q_n^0(x,x_0)-Q_n^0(x,-x_0)
\end{equation}
To take into account that the pair is absorved at point $x$, we have to exclude random paths where $x$ is visited more than once. Let the generating function for the probabilities $Q_n^W(x,x_0)$ be
\begin{equation}\label{def1}
\mathcal{P}(z)=\sum_{n=1}^{\infty} Q_n^W(x,x_0)
z^n=\sum_{n=1}^{\infty} p_n z^n 
\end{equation}
and consider the generating function of the probabilities $Q_n^W(x,x)$
\begin{equation}\label{def3}
\mathcal{R}(z)=\sum_{n=1}^{\infty} Q_n^W(x,x) z^n=\sum_{n=1}^{\infty}
r_n z^n 
\end{equation}
The probability we have after $f_n$ obeys the relation
\begin{equation}
p_n=f_n+f_{n-1}r_1+f_{n-2}r_2+\cdots+f_1r_{n-1}
\end{equation}
or, in terms of generating functions,
\begin{equation}
\mathcal{F}(z)=\frac{\mathcal{P}(z)}{\mathcal{R}(z)}
\label{relation}
\end{equation}
Finally,
\begin{equation}
\mathcal{P}(z)=\frac{e^{-x\sqrt{2y}}}{\sqrt{2y}}[e^{x_0\sqrt{2y}}-e^{-x_0\sqrt{2y}}],
\end{equation}
\begin{equation}
\mathcal{R}(z)=\frac{1}{\sqrt{2y}}[1-e^{-2\sqrt{2y}}],
\end{equation}
where we have taken $z=e^{-y}$, and the condition $z<1$ is needed to perform the integrations. From these two results and eq.\ref{relation},
\begin{equation}
\mathcal{F}(z)=\frac{\mathrm{sinh}x_0\sqrt{2y}}{\mathrm{sinh}x\sqrt{2y}}
\end{equation}
This generating function evaluated at the particular point\footnote{$z=1^-$ means that we must approximate $z=1$ from below {\it i.e.} $\mathrm{lim}_{\epsilon\to 0}\mathcal{F}(1-|\epsilon|)$.} $z=1^-$ gives us the desired probability
\begin{equation}
\sum_{n=1}^\infty f_n=\mathcal{F}(1^-)=\frac{x_0}{x}.
\end{equation}

\pagebreak

\end{document}